\documentclass[aps,prl,twocolumn,superscriptaddress]{revtex4}%
\usepackage{amsfonts}
\usepackage{amsmath}
\usepackage{amssymb}
\usepackage{subfig}
\usepackage{graphicx}%
\begin{document}
\title{Anisotropic scattering rate in Fe substituted Bi2212.}

\author{ M. Naamneh}
\affiliation{Physics Department, Technion-Israel Institute of Technology, Haifa 32000, Israel} 
\author{ Y. Lubashevsky}
\affiliation{Physics Department, Technion-Israel Institute of Technology, Haifa 32000, Israel} 
\author{E. Lahoud}
\affiliation{Physics Department, Technion-Israel Institute of Technology, Haifa 32000, Israel} 
\author{G. D. Gu}
\affiliation{Condensed Matter Physics and Materials Science Department, Brookhaven National Laboratory, USA}
\author{A. Kanigel}
\affiliation{Physics Department, Technion-Israel Institute of Technology, Haifa 32000, Israel}

\begin{abstract}
We measured the electronic structure of Fe substituted Bi2212 using Angle Resolved Photoemission Spectroscopy (ARPES). We find that the substitution does not change the momentum dependence  of the superconducting gap but induces a very anisotropic enhancement of the scattering rate. A comparison of the effect of Fe substitution to that of Zn substitution suggests that the Fe reduces T$_c$ so effectively because 
it supresses very strongly the coherence weight around the anti-nodes.  
\end{abstract}

\maketitle

Impurity substitution for Cu in the CuO planes of the cuprates was among the first evidence for the d-wave symmetry of the order parameter is these materials. S-wave superconductors are immune to non-magnetic impurities but are sensitive to magnetic impurities \cite{Anderson}. The way the magnetic impurities reduces T$_c$ in S-wave superconductors is well understood, 
in the cuprates, on the other hand, it was found that both magnetic and non-magnetic impurities reduces T$_c$ very effectively \cite{Alloul}. In fact, Zn, which is not magnetic, reduces T$_c$ faster than magnetic Ni. It is not clear how the impurities reduce T$_c$ in the cuprates, but it seems that the Abrikosov-Gorkov  pair-breaking mechanism \cite{AG} can not explain some aspects of the avialble experimental data. 

In order to study the effect of substitutional impurities on the electronic structure of the cuprates 
we performed Angle Resolved Photoemission Spectroscopy (ARPES)  measurements on 4 different $\rm Bi_2Sr_2Ca(Cu_{1-x}Fe_x)_2O_{8+\delta}$ (Fe-Bi2212) single crystals with x between zero and 3\%. The crystals were grown using a floating-zone furnace and were optimally doped.

\begin{figure}
\includegraphics[scale=0.35]{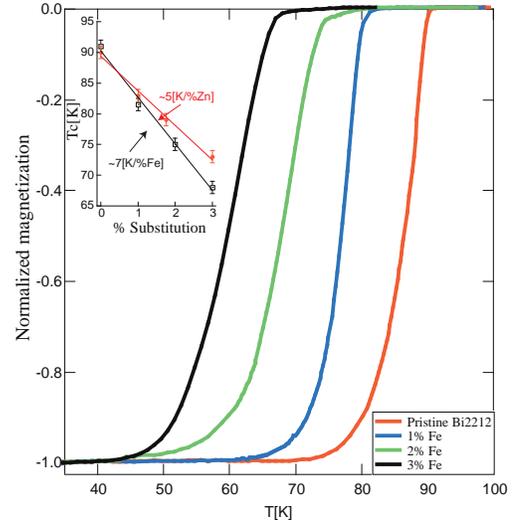}
\caption{Normalized magnetization as function of temperature for a pristine samples and the three Fe substituted samples. Inset: T$_c$ as function of the impurity concentration for Fe and Zn substituted Bi2212.}
\label{Fig1}
\end{figure}

In Fig. \ref{Fig1} we show the temperature dependence of the magnetization for the different samples, pristine Bi2212, and Fe-Bi2212 with 1\%, 2\% ad 3\% Fe substitution. 
As expected Fe substitution reduces T$_c$. In the inset we show T$_c$ as function of the substitution percentage; for these samples and for a set of Zn substituted Bi2212 films \cite{Zn_PRL}. $d$T$_c$/$d$(Impurity \%) is about -7.5K for Fe and about -5K for Zn.

\begin{figure}
\includegraphics[scale=0.35]{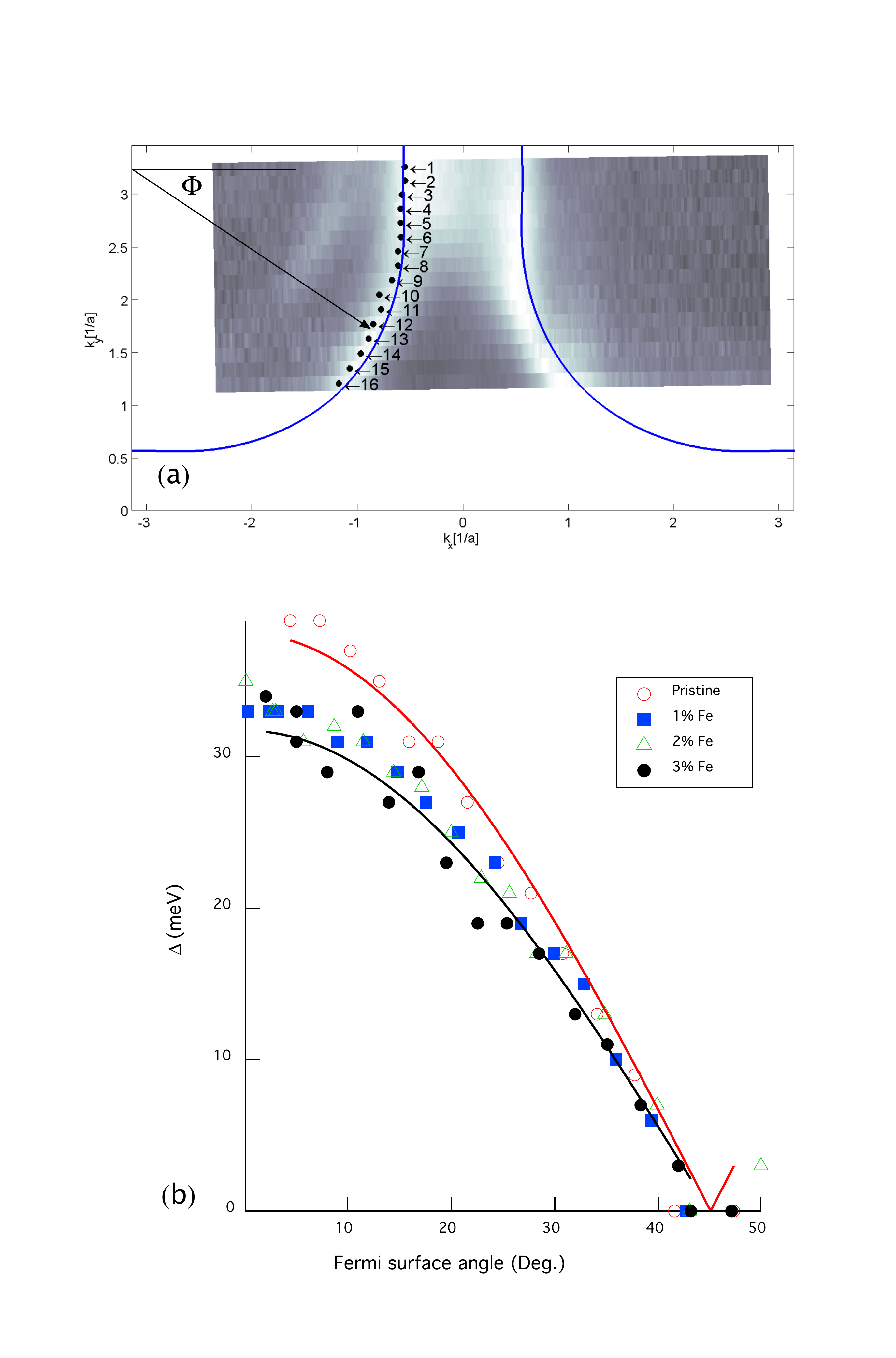}
\caption{(a) Intensity map for the 1\% Fe substituted sample around the Fermi-level. The intensity was integrated over 40meV around the Fermi-level to show the entire Fermi-surface. The solid-line represents a tight-binding model of the Fermi-surface. (b) The superconducting gap of a pristine samples and the three Fe substituted samples plotted as function of the Fermi-surface angle. The Fermi-surface angle is defined in Fig. \ref{Fig5}(a).}
\label{fig2}
\end{figure}

We used ARPES to measure both the superconducting gap and the momentum dependence of the scattering rate of the quasiparticles. The experiments were done at the PGM beam-line at the SRC, Madison WI. All data was obtained using a Scienta R4000 analyzer at a photon energy of 22eV. The energy-resolution was set to 12meV and the angular resolution to 0.2 Deg. All samples were cleaved in-situ at a pressure lower than 1 $\times$ 10$^{-10}$ Torr and at a temperature of 20K and measured at the same temperature. To measure the electronic structure we took several momentum-cuts parallel to the $\Gamma$-M direction, the momentum range we measured covers the Fermi-surface from node to anti-node. In Fig \ref{fig2}a we show as an example the cuts for the 1\% sample. The data shown was integrated over 40meV around the Fermi level.   
 
For each momentum-cut we follow the peak position in the Energy Distribution Curve (EDC). k$_F$ is defined as the point at which the peak is at minimal binding energy with respect to the Fermi level. The peak position of the EDC at k$_F$ after division by a resolution-broaden Fermi function defines the superconducting gap at this momentum. The Fermi level is found by measuring the density of states of a gold layer evaporated on the sample holder.

In Fig. \ref{fig2}b we show the gap as function of the Fermi-surface angle for all the samples at T=20K, the Fermi-surface angle is defined in Fig. \ref{fig2}a. Clearly, the momentum dependence of the gap is not changed. For all the samples we find a well defined nodal point where the gap vanishes.   This result is in agreement with recent electronic specific heat measurements in Ni substituted La$_{1-x}$Sr$_x$CuO$_4$ (LSCO) where it was shown that the magnetic field dependence of the specific heat follows the Volovik relation, a clear indication of a SC gap with line nodes \cite{Kurosawa}. 

This is also in agreement with our previous work where we have shown that the gap in Zn substituted Bi2212 is completely insensitive to the presence of the impurities \cite{Zn_PRL}, even around the node where the gap is small.  In the present case, it seems that the gap of all the substituted samples is identical but slightly smaller than the gap of the pristine sample.
This is due to a small difference in the doping level between the pristine and substituted samples. 

The insensitivity of the gap to the Fe content demonstrates again that in the cuprates the transition temperature is not set by the gap size, as T$_c$ of these samples varies by about 25\% but their superconducting gap is identical.  

Next we would like to examine the momentum dependence of the enhancement in the scattering rate induced by the Fe substitution. Assuming that Matthiessen's rule holds, the scattering rate due to the Fe impurities should be added to the scattering rate due to all other sources of scattering present in the pristine samples. Here we face a problem since in the normal state of Bi2212, ARPES finds very broad EDCs. Comparing EDC widths in prisitne and substituted samples above T$_c$ seems like an hopeless task since the impurities contribution to the total scattering rate is expected to be small. On the other hand, below T$_c$ sharp peaks are observed even in underdoped samples \cite{QP_PRL} and the impurities effect should be clear. We note that even below T$_c$ the width of the coherence peaks at the gap edge is mainly controlled by the single particle life-time \cite{phenomenology_PRB}. 

\begin{figure*}
\includegraphics[scale=0.45]{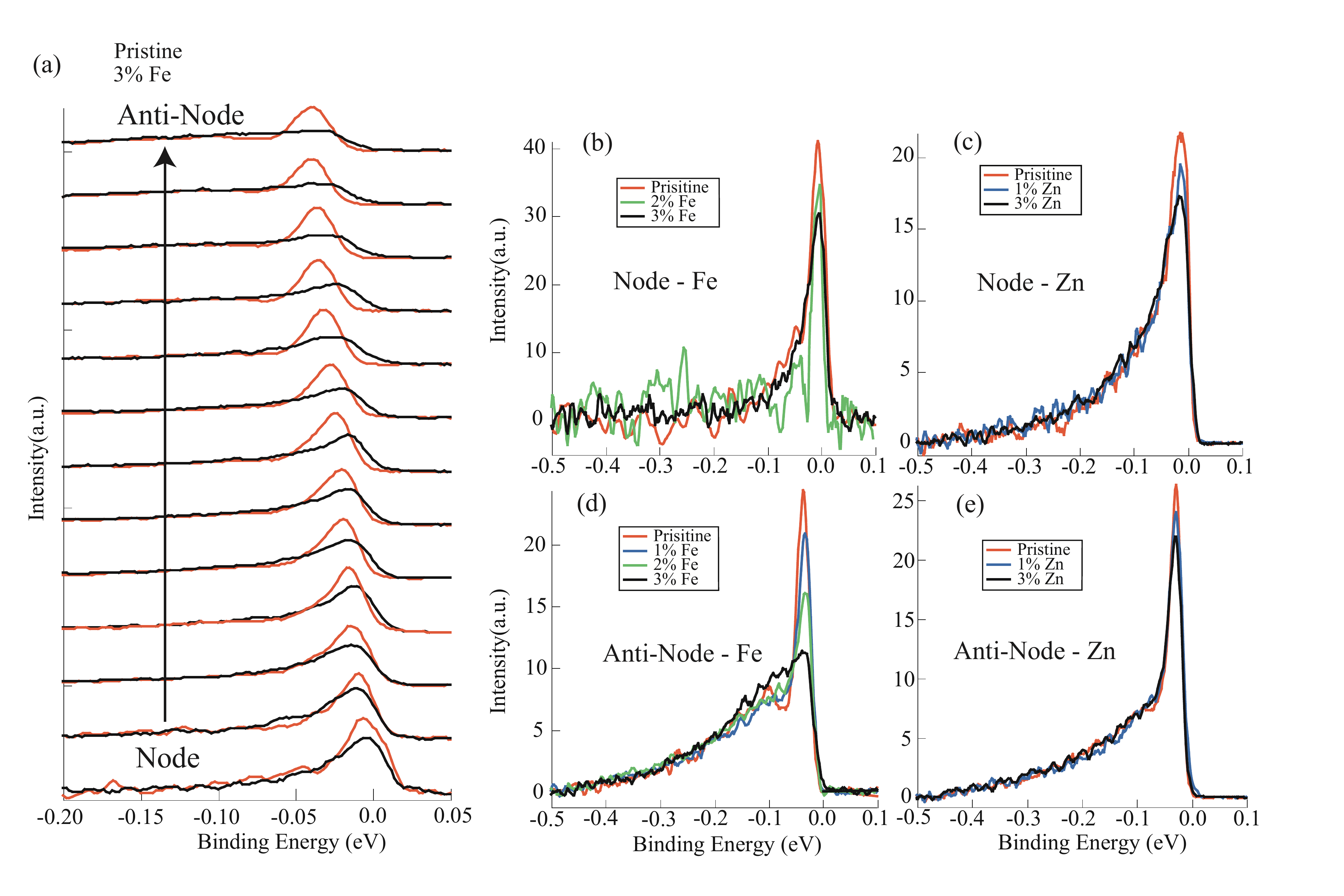}
\caption{Effect of the impurity substitution on the coherence peaks. (a) Comparison of the EDCs at k$_F$ of the 3\% Fe substituted sample and the pristine sample at 13 different k$_F$ points. (b) Effect of Fe substitution on the coherence peak at the Node. (c)  Effect of Zn substitution on the coherence peak at the Node. (d) Effect of Fe substitution on the coherence peak at the Anti-Node. (e) Effect of Zn substitution on the coherence peak at the Anti-Node. }
\label{Fig3}
\end{figure*}

We start by comparing the different EDCs without trying to extract the scattering rates.
In Fig. \ref{Fig3}a we compare EDCs for the pristine and the 3\% Fe samples at k$_F$ for 13 different cuts covering a region that extends from the node to the anti-node.  The EDCs were normalized to their high binding energy values and the background was removed by subtracting an EDC measured deep in the unoccupied region. Clearly the impurities reduce the spectral weight of the coherence peaks. The effect depends strongly on the momentum, near the anti-node the peak is almost gone completely. However, in the nodal region where the gap vanishes, the suppression is much weaker.

We found a different behavior in the Zn substituted samples, in that case we found that the change in scattering rate due to the impurities is isotropic \cite{Zn_PRL}.  To demonstrate how different is the effect of Zn and Fe we show a comparison of the nodal and anti-nodal EDCs for both in Fig \ref{Fig3}(b-e).  In the Zn substituted samples the peak-height is reduced as the substitution level is increased in a similar manner both at the node (Fig \ref{Fig3}b) and at the anti-node (Fig \ref{Fig3}c). While the Fe effect is similar around the node (Fig \ref{Fig3}d), the anti-nodal peak is completely washed-out by the Fe (Fig \ref{Fig3}e) indicating a large increase in the scattering-rate at that momentum region.

\begin{figure}[h!]
\includegraphics[width=8.5cm]{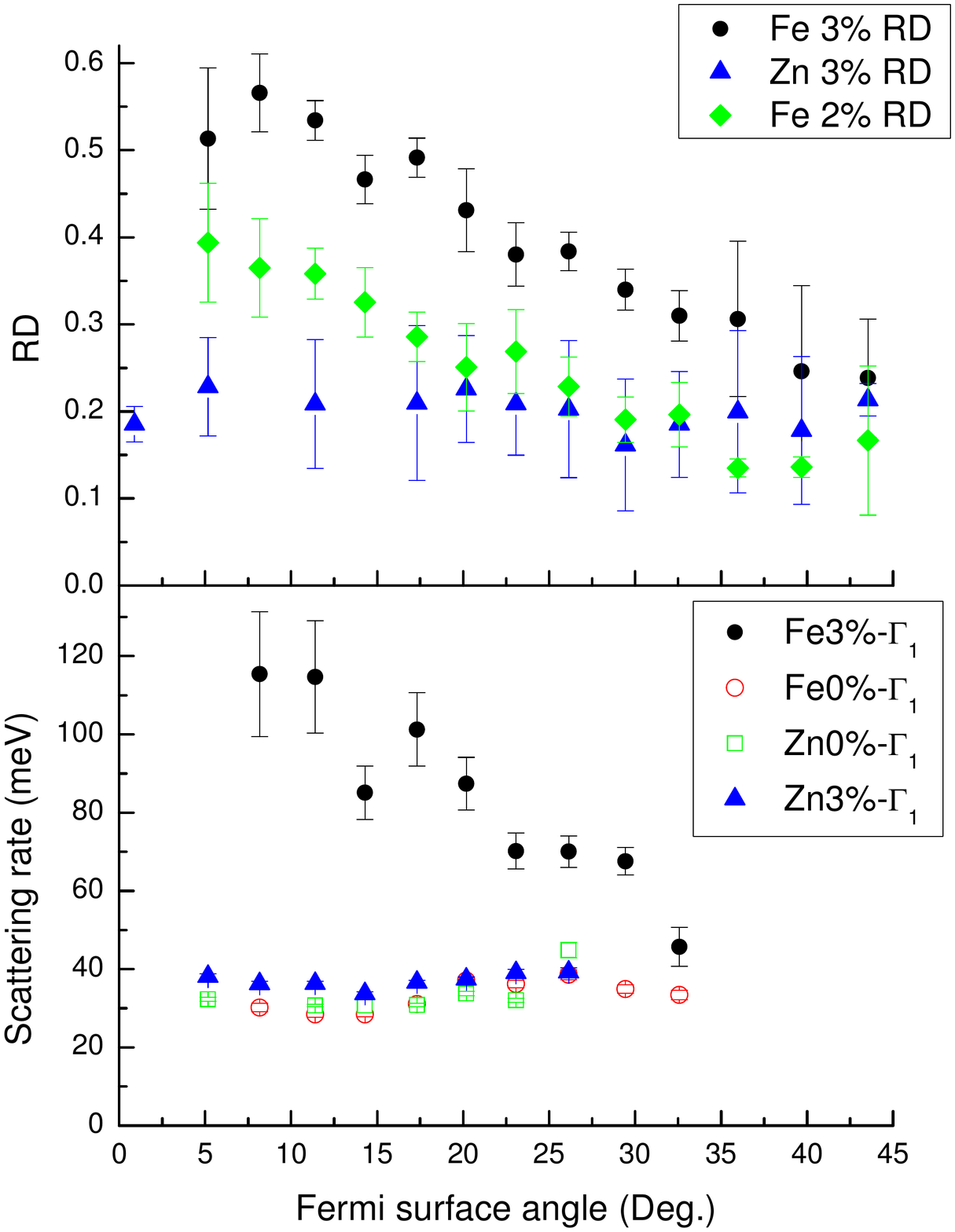}
\caption{Anisotropic scattering. (a) Relative difference between the coherence peak height of the substituted sample compared to the pristine sample, for for a 2\% and 3\% Fe substituted sample and a 3\% Zn substituted sample as function of the Fermi-surface angle. (b) Single-particle scattering rate, $\Gamma_1$ as function of the Fermi-surfce angle. We show data for a pristine and 3\% Fe substitution  crystals and a pristine and 3\% Zn substitution films. }
\label{Fig4}
\end{figure}

\begin{figure}[t]
\includegraphics[width=9cm]{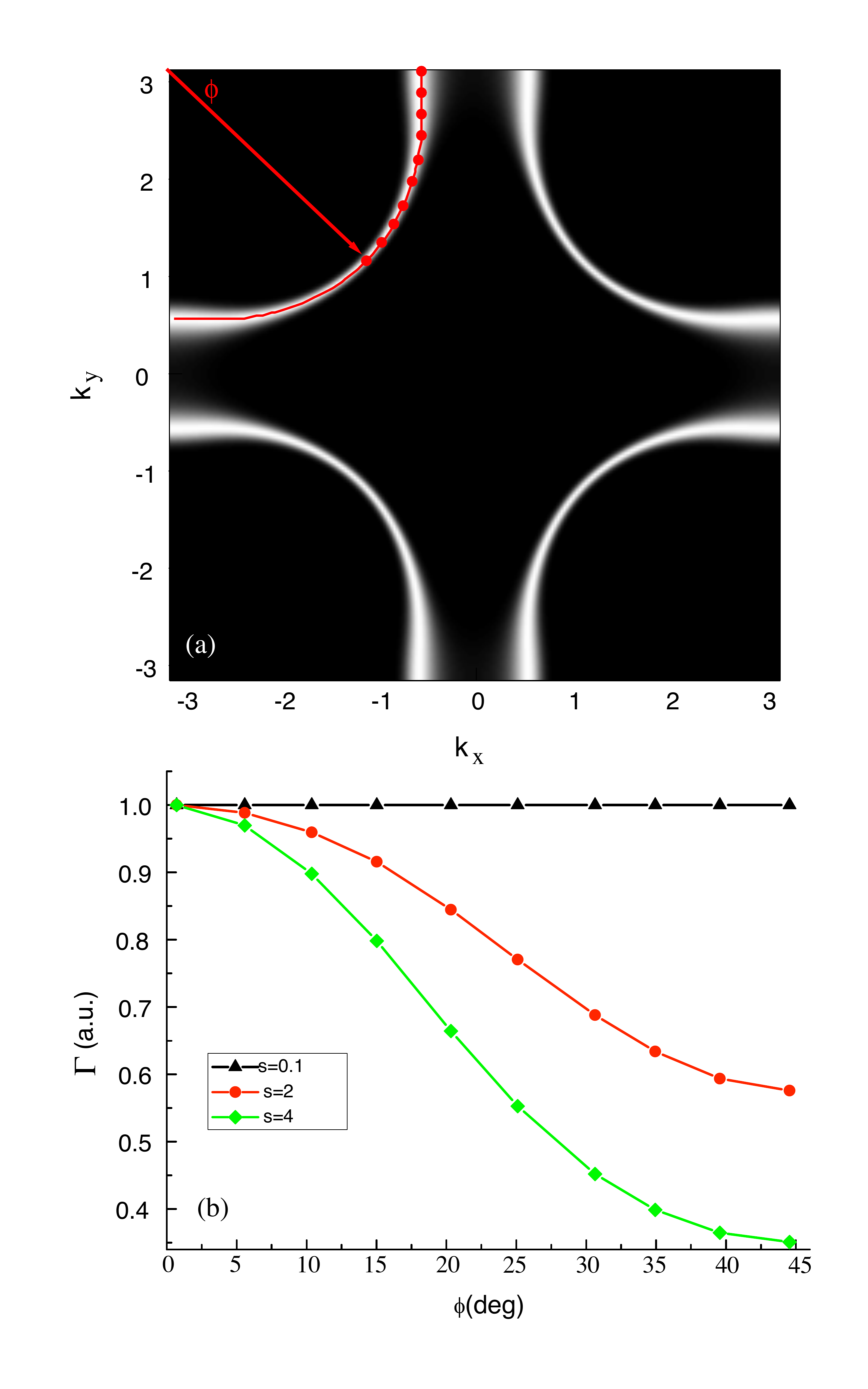}
\caption{Effect of the scattering range on the anisotropy of the scattering rate. (a) Shape of the Fermi-surface used in Eq. \ref{tau2}, we used a the tight-binding model of Ref \cite{Mike_TB} and 10 meV broadening. The red dots represents the points for which we show the scattering rate in panel (b). $\phi$ is the Fermi-surface angle. (b) Scattering rate as a function of the Fermi-surface angle for a Gaussian scattering-potential for various scattering-lengths. S represents the FWHM of the Gaussian in units of lattice constants. }
\label{Fig5}
\end{figure}

To get a more quantitative estimate of the influence of the two type of impurities on the coherence peaks, we plot in Fig.~\ref{Fig4}(a) the relative difference between the coherence peak height of the pristine sample and the impurity substituted sample. We define the relative difference as follows:
\begin{equation}
RD(\theta)=\frac{I(\theta)_{pristine}-I(\theta)_{impurity}}{I(\theta)_{pristine}}
\label{eqn:RD}
\end{equation}
 where $\theta$ is the Fermi-surface angle defined in Fig. \ref{Fig5}a and $I(\theta)$ is the intensity-maximum of the EDC at $k_F$ at this angle. Clearly, the Fe substitution produces an anisotropic increase in the scattering rate compared to the flat curve in the case of Zn substitution.
 
A similar conclusion can be drawn from a fit to the data using a BCS-line shape. We take a modified BCS self-energy of the form given in Ref \cite{phenomenology_PRB}:
\begin{equation} 
\Sigma(k,\omega)=-i\Gamma_1+\Delta(k)^2 / [\omega+ \varepsilon(k)+i\Gamma_0]
\label{spectral}
\end{equation}
where $\Gamma_1$ is the single-particle scattering rate, $\Delta(k)$ is the superconducting gap and $\varepsilon(k)$ is the electronic-dispersion.
To take into account pair-breaking we also include a second scattering rate, $\Gamma_0$, that should be thought of as the inverse life-time of the Cooper-pairs. Using Eq. \ref{spectral} we calculate the spectral function, convolve it with a Gaussian representing the experimental resolution, and fit it to the symmetrized EDCs.The $\Gamma_1$ values extracted from the fits are shown in Fig. \ref{Fig4}(b). We are showing the results for the pristine samples and for 3\% Zn and Fe substituted samples. 
The $\Gamma$-M orientation of the sample leads to an artificial broadening of the data around the node due to the steep dispersion, so we limit our fits to a Fermi-surface angle range of 0-30$^o$.  
Again, we find that the Zn substitution has an isotropic effect, but the Fe leads to an anisotropic increase in the single-particle scattering rate. Around the anti-node the scattering rate in the 3\% Fe sample is three times larger than the scattering rate found in the pristine sample.
We find that both Zn and Fe substitution increases the pair-breaking in agreement with previous work \cite{Dessau_Fe}, but this is a  small effect compared to the increase in the scattering-rate. 

The momentum dependent scattering rate of a metal is given by:
\begin{equation}
\frac{1}{\tau(k)}=\int \frac{dk'}{(2\pi)^3} W_{k,k'}[1-g(k')]
\label{tau1}
\end{equation}
where $W_{k,k'}$ is the scattering probability and $g(k)$ is the distribution function \cite{ashcroft}.
One can use the "Golden Rule" to calculate $W_{k,k'}$:
$$W_{k,k'}=\frac{2\pi}{\hbar} n_i \delta(\varepsilon(k) - \varepsilon(k')) |<k|U|k'>|^2$$ 
where $n_i$ is the impurity concentration and $U(r)$ is the scattering potential of a single impurity. 
Using the Fourier-transform of $U(r)$, we can use Eq. \ref{tau1} to calculate the scattering rate for quasiparticles on the Fermi-surface:
\begin{equation}
\frac{1}{\tau(k_F)} = \sum_{q} |U_{k-q}|^2 g(q) \delta(\varepsilon_F)
\label{tau2}
\end{equation}  
Eq. \ref{tau2} is a convolution between, $|U_k|^2$, the scattering potential in momentum-space squared and, $g(k)\delta(\varepsilon_F)$, a function that represents the shape of the Fermi-surface in momentum-space. 
For an impurity that creates a local potential, $U_k$ would be  constant and the scattering rate induced by these impurities would be completely isotropic. 
If the scattering potential varies in real-space then the momentum dependence  of the scattering rate can be non-trivial and has to be calculated.  

There is no way to measure directly the real-space structure of the scattering-potential, furthermore, calculations of the scattering potential induced by Zn and Ni substitution show that the real potential scattering depends strongly on the details of the impurity atom \cite{panagopolus,Hirscfeld_theory}. Our results suggests that by measuring the angular dependence of the enhancement in the single-particle scattering rate one could get some information about the scattering potential. 

The simplest way to reproduce qualitatively our results is to assume that the scattering potential has a finite range.  In Fig. \ref{Fig5} we show the results of solving Eq. \ref{tau1} for a Gaussian scattering-potential with different scattering ranges. 
  In  panel (a)  we show a function representing the momentum-distribution of electrons on the Fermi-surface calculated using a tight-binding model \cite{Mike_TB}.  In panel (b) we show the momentum dependence of the scattering-rate for different scattering ranges, the scattering-rate is plotted as a function of the Fermi-surface angle and should be compared to the data shown in Fig. \ref{Fig4}. 

The effect of impurity substitution on the local density of states of Bi2212 was studied extensively using Scanning Tunneling Microscopy (STM) \cite{Davis_MatToday}. In the presence of in-plane impurities sharp peaks in the density of states are found in the vicinity of the impurities. This peaks are resonances created in the superconducting gap and can be used to learn about the scattering potential. The energy of the resonance is given by $\Omega= - \frac{\Delta_0}{2 N_0 U_0 ln|8N_0U_0|}$ \cite{balatsky_review} where $N_0$  is the normal-state density of states and $U_0$ is the scattering potential. The stronger the scattering potential is the closer the resonance peak is to zero energy. 

For the case of non-magnetic Zn substituted Bi2212 it was found that the resonance is at -1.5meV and that the intensity of the peak in the density-of-states decays exponentially as one moves away from the impurity site \cite{Zn_STM}.  Ni impurities are found to be  weaker scatterers, based on the resonances found at $\pm$9meV and $\pm$19meV. Interestingly, it was found that the spectra around the Ni impurities conserve over-all particle-hole symmetry \cite{Ni_STM}. Zn, which is not magnetic, is found to suppress T$_c$ faster than Ni which is magnetic, this is in agreement with the STM data that shows that Zn is a stronger scatterer. 

Recently, STM data from Fe-Bi2212 samples became available, surprisingly Fe is found to be a weak scatterer, even weaker than Ni \cite{Boyer_APS}. This raises the question: how can such a weak scatterer reduce T$_c$ so effectively?  Our data suggests a possible explanation, Fe affects very efficiently the anti-nodal region of Bi2212. A similar mechanism was suggested to explain the effect of out-of-plane disorder. Out-of-plane disorder is known to reduce T$_c$ faster than in-plane-disorder for the same level of increase in the residual resistivity \cite{Fujita_PRL05}. Impurities away from the CuO planes induce a relatively smooth potential, so the effective scattering range should be long. As a result, the out-of-plane impurities will induce mainly forward scattering which will strongly reduce the life-time of the anti-nodal quasi-particles without increasing the residual resistivity \cite{Forward_scattering}.

Optical conductivity measurements using the same kind of samples found that the Fe substitution reduces the superfluid density of the samples \cite{OpticalCon}, this is expected to reduce T$_c$ since according to the Uemura relation T$_c$ $\propto$ n$_s$. It was shown \cite{Feng} that the anti-nodal quasi-particle spectral weight in Bi2212 is proportional to the superfluid density. This results is not well understood but it  suggests that the  Fe suppress T$_c$ so effectively by selectively reducing the spectral weight at the anti-node where it is maximal. 

In summary, we compared the effect of Zn and Fe substitution for Cu in the CuO planes of Bi2212. We find that both impurities have no effect of the superconducting gap. Both Zn and Fe reduce the quasi-particles life-time, but while the Zn effect is isotropic the Fe affects very strongly  the anti-nodal quasi-particles. We suggest that this is a result of the scattering range of the impurities.

We acknowledge useful discussions with Daniel Podolsky, Sasha Balatsky and Snir Gazit. 
This research was supported by the Israeli Science Foundation.  
 The Synchrotron Radiation Center is supported by NSF DMR-0084402. Work at Brookhaven is supported by the Office of Basic Energy Sciences, Division of Materials Sciences and Engineering, U.S. Department of Energy under Contract No. DE-SC00112704.

\end{document}